\documentstyle[aaspp4,11pt,flushrt]{article}

\def\ifm#1{\relax\ifmmode#1\else$\mathsurround=0pt #1$\fi}
\def\kms{~{\rm km\ s\ifm{^{-1}}}}

\def\mpc{~{\rm Mpc}}

\def\refpar{\par\hangindent=3em\hangafter=1}

\newcommand{\etal}{et al.}
\newcommand{\lya}{\mbox{${\rm Ly}\alpha$}}
\newcommand{\lyb}{\mbox{${\rm Ly}\beta$}}

\begin{document}

\title{\Large\bf Star-forming galaxies at very high redshifts}

\author{\large\bf Kenneth M. Lanzetta\altaffilmark{1},
Amos Yahil\altaffilmark{1} \& Alberto Fern\'andez-Soto\altaffilmark{2,3}}

{\small \noindent $^1$Astronomy Program, Department of Earth and Space
Sciences, State University of New York at Stony Brook, Stony Brook, NY
11794--2100, USA}

{\small \noindent $^2$Instituto de F{\'\i}sica de Cantabria, Consejo Superior
de Investigaciones Cient{\'\i}ficas, $^3$Departamento de F{\'\i}sica Moderna,
Universidad de Cantabria, Facultad de Ciencias, 39005 Santander, SPAIN}

\vspace*{0.1in}
\hrule
{\bf \noindent Analysis of the deepest available images of the sky, obtained by
the Hubble Space Telescope, reveals a large number of candidate high-redshift
galaxies.  A catalogue of 1,683 objects is presented, with estimated redshifts
ranging from $z=0$ to $z>6$.  The high-redshift objects are interpreted as
regions of star formation associated with the progenitors of present-day normal
galaxies at epochs reaching to 95\% of the time to the Big Bang.}
\vspace*{0.05in}
\hrule
\vspace*{0.1in}

  \noindent The longstanding effort to identify normal galaxies at high
redshifts has undergone dramatic progress in recent months.  New observations
by Steidel \etal$^1$ to magnitude $AB(6930)<25$ (where $AB(\lambda)$ is the
monochromatic magnitude at wavelength $\lambda$) have revealed a population of
galaxies at redshift $z \approx 3$ and have demonstrated that the Lyman-limit
spectral discontinuity and \lya-forest spectral decrement, which arise owing to
photoelectric absorption by neutral hydrogen along the line of sight, together
constitute the most prominent spectral signature of very distant galaxies.
This result has two important implications.  First, it provides a means of
identifying high-redshift galaxies.  The spectra of high-redshift galaxies are
characterized by (1) a complete absence of flux below the Lyman limit and (2)
strongly absorbed flux in the \lya\ forest.  This spectral signature is
observable by means of broad-band photometry and must apply irrespective of the
underlying spectral properties of the galaxies because it is imprinted by
intervening rather than intrinsic material.  Second, it allows high-redshift
interpretations of low-redshifts galaxies to be excluded.  The rate of
incidence of high-redshift Lyman-limit and \lya-forest absorbers is
sufficiently large that stochastic variations between different lines of sight
are essentially negligible.  High-redshift galaxies must exhibit the spectral
signature of Lyman-limit and \lya-forest absorption, and any observation to the
contrary is sufficient to rule out the possibility of a high redshift.

  The Hubble Deep Field (HDF) images, obtained by the Hubble Space Telescope
(HST) in December 1995, permit a search for the spectral signature of
Lyman-limit and \lya-forest absorption to magnitudes far fainter than were
previously accessible.  The images were obtained with the Wide Field Planetary
Camera 2 (WFPC2) through four filters spanning near-ultraviolet through
near-infrared wavelengths to a roughly uniform detection limit approaching $AB
= 30$.  The images are in principle sensitive to galaxies at redshifts as large
as $z \approx 7$, beyond which the Lyman limit is redshifted past the response
of the WFPC2 filters.

  We have estimated redshifts of a large number of galaxies in the HDF images
by fitting galaxy spectral energy distributions, including the effects of
intervening Lyman-limit and \lya-forest absorption, to precise photometry of
objects detected in the images.  Here we describe the analysis and present a
catalogue of 1,683 objects of magnitude $AB(8140)\la 30$ at estimated redshifts
ranging from $z=0$ to $z>6$.  The catalogue is essentially complete for
magnitudes $AB(8140)<28$ (1,104 objects), at which 367 objects are at estimated
redshift $z=0-1$, 512 objects are at $z=1-2$, 135 objects are at $z=2-3$, 54
objects are at $z=3-4$, 30 objects are at $z=4-5$, two objects are at $z=5-6$,
and four objects are at $z>6$.  Even if the high-redshift identifications are
incorrect, we show by simulations that most real galaxies of those magnitudes
and redshifts would be detected and hence that the surface densities we find
represent strict upper limits to the actual surface densities.  The rapid
decline in the number of objects at estimated redshift $z>2$ is therefore
significant and tightly constrains models of galaxy formation and evolution
(A.Y., K.M.L., A.\ Campos, and A.F.-S., manuscript in preparation).

  If the estimated redshifts are approximately correct, then the high-redshift
objects typically have ultraviolet luminosities $\sim 10^9 - 10^{10}$ times the
solar luminosity $L_\odot$, sizes $\sim 1$ kpc, and co-moving spatial densities
that vary between $0.05$ and $0.01$ Mpc$^{-3}$ at redshifts between $z=2.5$ and
$z=6$.  The ultraviolet luminosities and sizes are similar to those of nearby
starbursting galaxies, while the co-moving spatial densities are comparable to
that of present-day galaxies.  (Throughout we adopt a Hubble constant
$H_0=100\kms\mpc^{-1}$ and an Einstein--de Sitter cosmology with $\Omega=1$ and
$\Lambda=0$.)  We interpret these objects as galactic or proto-galactic regions
of star formation associated with the progenitors of present-day normal
galaxies.

\section*{\large\bf Object detection}

  The first goal of the analysis is to detect objects in the F814W images and
to measure their spectral energy distributions in the F300W, F450W, F606W, and
F814W images.  (The spectral sensitivities of these images peak at roughly 3000
\AA, 4500 \AA, 6060 \AA, and 8140\AA, respectively.)

  First, we obtained and processed the HDF images (produced by the Version 2
drizzle algorithm) made available to the HST archive on February 29, 1996.  We
considered only the Wide Field Camera images, which are significantly deeper
than the Planetary Camera images.  We trimmed the images to include only
columns and rows 191 through 1970 (because the images are of inferior quality
at the edges) and eliminated deviant pixels by setting the value of any pixel
that differed by more than $3 \sigma$ from a local $3 \times 3$ pixel median
equal to that median (because the images contain a number of single-pixel
positive noise spikes or hot pixels).  The angular extent of each trimmed image
is $71.2 \times 71.2$ arcsec$^2$, and the total angular area covered by the
trimmed images is $4.22$ arcmin$^2$.

  Next, we detected objects in the F814W images using the SExtract program$^2$.
The object detection criteria can be set in different ways for different
purposes, and we adopted a conservative approach with the aim of achieving high
reliability.  Specifically, we changed the default parameters of the program
by: (1) smoothing the images by a Gaussian filter of ${\rm FWHM} = 0.12$
arcsec (approximately the width of the point spread function) to aid the
detection of faint sources,$^3$ (2) adopting a detection threshold of at least
10 contiguous pixels with signal-to-noise ratio $>1.4$ each, and (3)
eliminating objects that are close to other objects by setting ${\rm
CLEAN\_PARAM} = 2.0$.  Criterion (2) is equivalent to a surface brightness
limit $\mu_{AB(8140)}<26.1$ arcsec$^{-2}$ over a minimum area of
0.016 arcsec$^2$, which corresponds to an isophotal limiting magnitude of
$AB(8140)=30.6$.  A total of 1,683 objects were detected, the faintest of which
is of total magnitude $AB(8140)=30.0$.

  Next, we tested the completeness of the catalogue by re-applying the
detection algorithm for different values of the smoothing and thresholding
parameters.  The object list does depend on these parameters, but for objects
of magnitude $AB(8140)<28$ we found less than 1\% variation for any plausible
choices of the parameters.  We conclude that the catalogue is essentially
complete to magnitude $AB(8140)< 28$.

  Next, we determined the local covariance and background surrounding each
object in the images.  We considered a square region of at least $41 \times 41$
pixels surrounding each object (larger regions for larger objects) and excluded
pixels associated with any object before measuring the local covariance and
background.  The local covariances must be used to determine accurate
photometric uncertainties because the drizzled HDF images are significantly
correlated over adjacent pixels ($\approx 40\%$ with immediate neighbors,
$\approx 10\%$ with diagonal neighbors, and less at larger separations).

  Last, we measured the spectral energy distribution of each object detected in
the images.  We used SExtract segmentation maps to identify the nonoverlapping
pixels associated with each object and measured the flux of each object in each
of the four images within its unique segmentation aperture.  The choice of
identical apertures in the four images assures that exactly the same portion of
each object is measured in each image, which is essential for establishing
meaningful spectral energy distributions.  We subtracted the local background
(which was generally negligible) and used the local $3 \times 3$ covariance
matrices (between each pixel and its immediate and diagonal neighbors) to
determine the photometric uncertainties.

\section*{\large\bf Redshift determination}

  The second goal of the analysis is to estimate redshifts of all objects by
comparing measured and modelled spectral energy distributions.

  First, we modelled spectral energy distributions of galaxies at redshifts
$0<z<7$.  We adopted the galaxy spectra (E/S0, Sbc, Scd, and Irr galaxies) of
Coleman, Wu, \& Weedman$^4$, extrapolating at wavelengths less than 1400 \AA\
using results of Kinney \etal$^5$.  We chose not to apply evolutionary
corrections to these spectra because such corrections are uncertain and because
the adopted galaxy spectra already span a wide range of spectral properties.
We incorporated the effects of intrinsic and intervening Lyman-limit absorption
by assuming that galaxies are optically thick to ionizing radiation.  This
assumption has recently been verified by Leitherer \etal$^6$ for nearby
star-forming galaxies, but in any case the mean free path for intervening
Lyman-limit absorption is sufficiently small$^8$ to make this assumption valid
at all but the lowest redshifts ($z \approx 2.3$) for which the Lyman limit is
of interest.  We accounted for intervening \lya\ and \lyb\ absorption by
applying measurements of Madau$^8$ and unpublished measurements of J.\ Webb of
the \lya-forest flux decrement parameters $D_A$ and $D_B$.  These measurements
extend over the redshift range $0<z<5$ and were extrapolated to $z=7$ using a
simple fit.  We integrated the redshifted spectra with the throughputs of the
F814W, F606W, F450W, and F300W filters (including system throughputs) to derive
the model spectral energy distributions.  Figure 1 shows the expected $AB$
magnitudes (of the four galaxy types) through the WFPC2 filters of a galaxy of
absolute magnitude $M_{AB(4500)}=-20$.  We also modelled the spectral energy
distribution of an M star using the spectrum of Jacoby, Hunter, \&
Christian$^{10}$, extrapolating at wavelengths longward of 7500~\AA\ using our
own observations.

  Next, we constructed redshift likelihood functions of all objects in the
images by comparing measured and modelled spectral energy distributions.
Assuming that flux uncertainties are normally distributed, the likelihood
$L(z,T)$ of obtaining measured fluxes $f_i$ with uncertainties $\sigma_i$ given
modelled fluxes $F_i(z,T)$ at an assumed redshift $z$ for spectral type $T$ and
normalization $A$ over the four filters $i=1-4$ is
\begin{equation}
L(z,T) = \prod_i \exp \left\{ -\frac{1}{2} \left[ \frac{f_i - A F_i(z,T)}
{\sigma_i} \right]^2 \right\}.
\end{equation}
For each object, we maximized Equation (1) with respect to spectral type and
normalization to determine the redshift likelihood function $L(z)$ and
maximized $L(z)$ with respect to redshift to determine the maximum-likelihood
redshift estimate.  We did not attempt to assign relative abundances to the
different spectral types as functions of redshift but simply gave all types
equal weight in the redshift likelihood functions.

  The result of the analysis is a redshift likelihood function and
maximum-likelihood redshift estimate of each of the 1,683 objects detected in
the images.  Spectral energy distributions and redshift likelihood functions of
a representative sample of the objects are shown in Fig.\ 2, and surface
densities of the objects as functions of redshift and limiting magnitude are
given in Table 1.  The complete catalogue of objects, as well as their
spectral energy distributions and redshift likelihood functions are abailable
as Supplementary Information, together with optimally added composite spectral
energy distributions of the objects.  The spectra of the objects fall into
distinct classes depending on the estimated redshift.  At redshift $z \la2.3$
the spectra exhibit no significant absorption by intervening material and are
similar to the redshifted spectra of present-day galaxies.  At redshift $2.5\la
z \la4$ the spectra are characterized by strong flux in the F814W and F606W
images, detectable flux in the F450W images, and no detectable flux in the
F300W images.  At redshift $4\la z \la5.5$ the spectra are characterized by
strong flux in the F814W images, detectable flux in the F606W images, and no
detectable flux in the F450W and F300W images.  Finally, at redshift $z \ga6$
the spectra are characterized by strong flux in the F814W images and no
detectable flux in the F606W, F450W and F300W images.

\section*{\large\bf Confirming and corroborating evidence}

  Several lines of evidence support the redshift estimates obtained by the
analysis.

  The redshift estimates are in good agreement with the spectroscopic redshifts
recently reported by Steidel \etal$^{10}$ for six high-redshift objects, by
Cowie$^{11}$ for 30 medium-redshift objects, and by Moustakas \etal$^{12}$ for
eight medium- and high-redshift objects, as is illustrated in Fig.\ 3.  Of the
44 objects, three high-redshift galaxies have redshifts underestimated by
$\Delta z \approx 1$ due to confusion with the UV emission of other faint
sources, and four bright medium-redshift objects, of magnitudes $AB(8140) =
20.5-22.8$, have redshifts overestimated by $\Delta z > 1$ because the
uncertainties used in Equation (1) do not include the cosmic variance.  The
estimated redshifts of the remaining 37 objects agree with the spectroscopic
ones with an rms difference $\Delta z=0.15$.

  The surface density of high-redshift objects found by Steidel \etal$^1$,
$0.40 \pm 0.07$ arcmin$^{-2}$ to magnitude $AB(6930)<25$ at redshift $3<z<3.5$,
is in good agreement with that found in our analysis, $1.2 \pm 0.5$
arcmin$^{-2}$.  Furthermore, the surface densities reported in Table 1 decrease
more or less monotonically with increasing redshift, which is in general accord
with what is expected for distant galaxies due to luminosity distance and
surface brightness effects but is not necessarily expected if the redshifts are
generally in error.

  We conducted two tests to verify that objects detected in the F814W image but
not in the other images (and hence identified by the analysis at estimated
redshift $z>6$) are real and not due to noise fluctuations.  First, we sought
to identify spurious objects in the ``negative'' images, which were formed by
reversing the sign of each pixel in the images.  Using exactly the same
procedures that were used on the positve images, we detected only three
spurious objects of magnitudes $AB(8140) = 28.5 - 29.6$ in the negative images,
of which only one is at ``estimated redshift'' $z>6$, compared with 16 objects
at $z>6$ in the positive images, of which one, four, and 13 are of magnitude
$AB(4500) < 27$, 28, and 29, respectively.  This rules out spurious detections
due to a symmetric noise distribution.  To eliminate the possibility of
spurious detections due to a skewed noise distribution, caused by noise added
to weak sources below the detection limit, we repeated the analysis with the
roles of the F814W and F450W images interchanged.  (These two images have
approximately the same limiting magnitude; this test cannot be conducted with
the F606W image, which is almost a magnitude deeper.)  As none of the spectra
are expected to peak at 4,500 \AA\ and be undetected at higher wavelengths, any
detection with ``estimated redshift'' $z>4$ would be spurious.  The highest
estimated redshift found in the test was, in fact, $z=3.64$.

  The spectra of the extreme objects at estimated redshift $z>6$ are consistent
with redshifted Lyman-limit absorption of high-redshift galaxies but
inconsistent with many other interpretations, including lower-redshift
galaxies, line-dominated galaxies, and heavily reddened lower-redshift
galaxies.  The brightest of these objects (object 1,668 in the full catalogue;
see Supplementary Information) shows a $3 \sigma$ lower limit of 12 to the
F814W/F606W flux ratio, and the composite spectral energy distribution of these
objects shows a $3 \sigma$ lower limit of 20 to the F814W/F606W flux ratio.  No
other astrophysical object of which we are aware satisfies these constraints.
Specifically, spectrophotometric atlases of galaxies$^{13,14}$ show no galaxy
with a spectral decrement anywhere near a factor of 20 over any wavelength
range that might be redshifted into the F814W and F606W images, and extreme
line-dominated galaxies usually exhibit strong ultraviolet continuum
radiation$^{15}$.

  The possibility that the objects might be heavily-reddened low-redshift
galaxies is unlikely on several accounts.  First, three magnitudes of
reddening, which corresponds to about 9 magnitudes of extinction in the F814W
images, would be required to suppress the F606W images by a factor of 20
relative to the F814W images (assuming a reasonably flat unobscured spectrum).
If it were heavily reddened, object 1,668 would have an unobscured magnitude of
$AB(8140)=26-9=17$, yet it is smaller than 1 arcsec.  Second, if the objects
were heavily reddened, then they would be very noticeable at longer
wavelengths.  Object 1,668 would be of magnitude $AB(19000) \approx 21-22$.
Infrared images of the HDF area of the sky have been checked (L. Cowie,
personal communication) and place an approximate, $1 \sigma$, limit of
$AB(19000)\ga25$ on this object, thus ruling out the possibility of heavy
reddening in this case.  Third, it is necessary to account for the surface
density of the objects, that is, $\sim 1$ arcmin$^{-2}$.  Odd, heavily
reddened, nearby galaxies cannot serve as prototypes for the galaxies at
estimated redshift $z>6$ unless the density of such local galaxies can be shown
to be high enough.

  Even if the high-redshift estimates are incorrect, the results of the
analysis can be used to set strict upper limits to the surface density of real
high-redshift galaxies because such galaxies must exhibit the spectral
signature of Lyman-limit and \lya-forest absorption and so must be identified
by the analysis as high-redshift objects.  Such galaxies would fail to be
identified by the analysis as high-redshift galaxies only if they were blocked
by bright, foreground galaxies or were incorrectly identified as low-redshift
galaxies due to photometric uncertainties.  To test both of these effects, we
placed ``synthetic'' models of all the objects of magnitude $AB(8140)<28$
(matching both the magnitudes and sizes of the real objects) at random
locations within the images and sought to recover the redshifts by exactly the
same analysis procedures used to identify the real galaxies.  Figure 4 shows
that the great majority of the objects are recovered at their input redshifts
with an r.m.s.\ difference $\Delta z=0.11$, similar to the redshift difference
found between the estimated and spectroscopic redshifts.  The remaining objects
have discordant estimated redshifts due to photometric uncertainties and
confusion with faint sources, but none have estimated redshifts $z>5$.  (Cosmic
variance is not an issue since the simulated objects were modelled with the
spectral energy distributions used in the analysis.)

  Many of the objects, including some of those with highest estimated
redshifts, are clearly spatially resolved, which excludes Galactic stars as
possibilities.  (Their spectra are, in any case, inconsistent with those of M
stars.)

  We therefore conclude that the redshifts are probably correct.

\section*{\large\bf Galaxies in the early universe}

\noindent If the redshifts we have estimated are approximately correct, then
the high-redshift objects typically have ultraviolet luminosities $\sim 10^9 -
10^{10} L_\odot$, sizes $\sim 1$ kpc, and co-moving spatial densities that vary
between $0.05$ and $0.01$ Mpc$^{-3}$ at redshifts between $z=2.5$ and $z=6$.
The luminosities and sizes are modest in comparison to luminous galaxies but
are similar to those of nearby starbursting galaxies$^{16}$.  On the other
hand, the co-moving spatial density of the objects is comparable to or even
larger than that of luminous galaxies.  The objects therefore appear to
represent star formation in small, concentrated regions rather than in
galaxy-sized objects.  At the early epochs spanned by the objects, the
dynamical timescale of galaxy-sized objects is comparable to the age of the
universe, which suggests that we may be witnessing the first star formation
associated with the initial collapse of galaxies.  In fact, the rapid decline
in surface density as a function of redshift for $z>2$ severely constrains
models for galaxy formation and evolution (A.Y., K.M.L., A.\ Campos, and
A.F.-S., manuscript in preparation).  In any event, we interpret the objects as
galactic or proto-galactic regions of star formation associated with the
progenitors of present-day normal galaxies at epochs reaching back 95\% of the
time to the Big Bang.

\vspace*{0.1in}
\hrule

\begin{small}
\noindent Received 7 March 1996; accepted 4 June 1996.

\noindent 1. Steidel, C., Giavalisco M., Pettini M., Dickinson M. \& Adelberger
K. Astrophys.\ J.\ in press, Los Alamos preprint astro-ph/9602024 (1996).

\noindent 2. Bertin, E. \& Arnouts, S. Astr.\ Astrophys.\ Suppl.\ in press
(1996).

\noindent 3. Irwin, M. J. Mon.\ Not.\ R. astr.\ Soc.\ {\bf 214}, 575--604
(1985).

\noindent 4. Coleman, G. D., Wu, C. C. \& Weedman, D. W. Astrophys.\ J. Suppl.\
{\bf 43}, 393--416 (1980).

\noindent 5. Kinney, A. L., Bohlin, R. C., Calzetti, D., Panagia, N. \& Wyse,
R. F. G. Astrophys.\ J. Suppl.\ {\bf 86}, 5--93 (1993).

\noindent 6. Leitherer, C., Ferguson, H. C., Heckman, T. M. \& Lowenthal, J. D.
Astrophys.\ J. {\bf 454}, L19--22 (1995).

\noindent 7. Lanzetta, K. M. Astrophys.\ J. {\bf 375}, 1--14 (1991).

\noindent 8. Madau, P. Astrophys.\ J. {\bf 441}, 18--27 (1995).

\noindent 9. Jacoby, G. H., Hunter, D. A. \& Christian C. A. Astrophys.\ J.
Suppl.\ {\bf 56}, 257--281 (1984).

\noindent 10. Steidel, C., Giavalisco M., Dickinson, M. \& Adelberger,
K. L. Astr.\ J. in press, Los Alamos preprint astro-ph/9604140 (1996).

\noindent 11. Cowie, L. L. http://www.ifa.hawaii.edu/~cowie/hdf.html (1996).

\noindent 12. Moustakas, L., Zepf, S., \& Davis, M.
http:////astro.berkeley.edu/davisgrp/HDF/ (1996).

\noindent 13. Kennicutt, R. C. Astrophys.\ J. Suppl.\ {\bf 79}, 255--284
(1992).

\noindent 14. Liu, C, T. \& Kennicutt, R. C. Astrophys.\ J. Suppl.\ {\bf 100},
325--346 (1995).

\noindent 15. Thuan, T. X. in ``Massive Stars in Starbursts'', eds.\ C.
Leitherer \etal\ (Cambridge: Cambridge University Press), pp. 183--203 (1991).

\noindent 16. Meurer, G. R., Heckman, T. M., Leitherer, C., Kinney, A., Robert,
C. \& Garnett, D. R. Astr.\ J. {\bf 110}, 2665--2691 (1995).
\end{small}

\acknowledgments

\noindent SUPPLEMENTARY INFORMATION.  Available on {\em Nature's} World-Wide
Web site http:\-//www.\-nature.\-com.  Paper copies are available from Mary
Sheehan at the London editorial office of {\em Nature}.

\noindent ACKNOWLEDGEMENTS.  We are grateful to X.\ Barcons, A.\ Burrows, A.\
Campos, L.\ Cowie, J.\ Lattimer, D.\ Peterson, R.\ Puetter, M.\ Simon, P.\
Solomon, F.\ Walter, and J.\ Webb for fruitful discussions; E.\ Bertin, L.\
Cowie, R.\ Rebolo, and J.\ Webb for software and data used in the analysis; S.\
Zoonematkermani for assisting with graphical display of the HDF images, and to
R.\ Williams and the staff of STScI for providing access to the reduced HDF
images.  This research was supported by NASA, the Dudley Observatory, a Spanish
MEC studentship, and the Spanish DGICYT.

\noindent CORRESPONDENCE should be addressed to K.M.L. (email:
lanzetta@\-sbastc.\-ess.\-sunysb.\-edu).

\newpage

\begin{center}
\begin{tabular}{p{1.0in}cccc}
\hline
\multicolumn{5}{c}{TABLE 1 Galaxy surface densities} \\
\hline
& \multicolumn{4}{c}{Surface Density (arcmin$^{-2}$)} \\
\cline{2-5}
\multicolumn{1}{c}{$z$} & $AB(8140)<25$ & $AB(8140)<26$ & $AB(8140)<27$ &
$AB(8140)<28$ \\
\hline
$0.0 - 0.5$ \dotfill &   7.1 (0.6) & 13.3 (0.9) & 22.0 (1.1) & 35.5 (1.4) \\
$0.5 - 1.0$ \dotfill &  20.4 (1.1) & 28.0 (1.3) & 39.1 (1.5) & 51.4 (1.7) \\
$1.0 - 1.5$ \dotfill &  13.0 (0.9) & 23.7 (1.2) & 42.7 (1.5) & 65.2 (1.9) \\
$1.5 - 2.0$ \dotfill &   8.3 (0.7) & 18.5 (1.0) & 34.6 (1.4) & 56.2 (1.8) \\
$2.0 - 2.5$ \dotfill &   2.1 (0.3) &  8.1 (0.7) & 15.2 (0.9) & 24.6 (1.2) \\
$2.5 - 3.0$ \dotfill &   0.9 (0.2) &  1.9 (0.3) &  3.1 (0.4) &  7.3 (0.6) \\
$3.0 - 3.5$ \dotfill &   1.2 (0.3) &  2.1 (0.3) &  4.5 (1.5) &  8.3 (0.7) \\
$3.5 - 4.0$ \dotfill &   0.2 (0.2) &  1.2 (0.3) &  2.6 (0.4) &  4.5 (0.5) \\
$4.0 - 4.5$ \dotfill &   0.2 (0.2) &  0.7 (0.2) &  2.8 (0.4) &  5.2 (0.5) \\
$4.5 - 5.0$ \dotfill &   0.2 (0.2) &  0.2 (0.2) &  0.5 (0.2) &  1.9 (0.5) \\
$5.0 - 5.5$ \dotfill &   0.0 (0.2) &  0.0 (0.2) &  0.2 (0.2) &  0.5 (0.3) \\
$5.5 - 6.0$ \dotfill &   0.0 (0.2) &  0.0 (0.2) &  0.0 (0.2) &  0.0 (0.2) \\
$> 6.0$     \dotfill &   0.0 (0.2) &  0.0 (0.2) &  0.2 (0.2) &  0.9 (0.2) \\
\hline
\end{tabular}
\end{center}

\newpage

\figcaption{Expected $AB$ magnitudes of the four galaxy types---E (solid), Sbc
(dotted), Scd (short dashed) and Irr (long dashed)---through the WFPC2 filters
for galaxies of absolute magnitude $M_{AB(4500)}=-20$ ($H_0=100\kms\mpc^{-1}$).
The top panels are for $\Omega=0$ and the bottom panels for $\Omega=1$.  The
WFPC2 images are sensitive to galaxies at redshifts as large as $z \approx 7$,
beyond which the Lyman limit is redshifted past the response of the F814W
filter.}

\figcaption{Sample spectral energy distributions (left panels) and redshift
likelihood functions (right panels) of 24 objects of magnitude $AB(8140)<28$.
For the spectral energy distributions, vertical error bars show uncertainties,
horizontal error bars show bin sizes, and solid curves show best-fit model
spectral energy distributions.  Rough uncertainties for the derived redshifts
can be estimated by identifying the redshift regions over which the redshift
likelihood functions exceed $\exp(-1/2)$ times the maximum-likelihood values.
However, the redshift likelihood functions are, in general, quite complicated
and often show multiple local maxima.  Moreover, they do not take into account
the cosmic variance.  For these reasons, the uncertainties determined by this
method provide only rough indications of the true uncertainties.}

\figcaption{Redshifts estimated in this paper versus spectroscopic measurements
by Steidel \etal$^{10}$ (solid circles), Cowie$^{11}$ (crosses), and Moustakas
\etal$^{11}$ (open boxes).  Of the 44 objects, three high-redshift galaxies
have redshifts underestimated by $\Delta z \approx 1$, owing to confusion with
the UV emission of other faint galaxies, and four bright ($AB(8140) =
20.5-22.8$) medium-redshift objects have redshifts overestimated by $\Delta z >
1$, because the uncertainties used in Equation (1) do not include the cosmic
variance.  The estimated redshifts of the remaining 37 objects agree with the
spectroscopic ones to an r.m.s.\ difference $\Delta z=0.15$.}

\figcaption{Output redshifts $z_{\rm out}$ versus input redshifts $z_{\rm in}$
of the synthetic models of the objects.  Of the 1,104 objects included into the
simulation, 170 (or 15\%) were by chance placed on or near other objects and
showed a brightening of $\Delta AB(8140)>0.15$.  The 934 remaining objects were
recovered with an r.m.s.\ difference $\Delta AB(8140)=0.045$ between the input
and output magnitudes.  Of these, 97 (or 9\% of the original sample) were
identified by the analysis with redshift differences $z_{\rm in} - z_{\rm
out} > 0.5$, with typical differences around 2.  The remaining 837 (or 76\% of
the original sample) were identified at output redshifts essentially equal to
the input redshifts, with an r.m.s.\ difference $\Delta z=0.10$, which is
comparable to that between the estimated redshifts and those measured by
Steidel \etal$^{10}$, Cowie$^{11}$, and Moustakas \etal$^{12}$.}

\end{document}


\title{\Large\bf Star-forming galaxies at very high redshifts \\
Supplementary information}

\author{\large\bf Kenneth M. Lanzetta\altaffilmark{1},
Amos Yahil\altaffilmark{1} \& Alberto Fern\'andez-Soto\altaffilmark{2,3}}

{\small \noindent $^1$Astronomy Program, Department of Earth and Space
Sciences, State University of New York at Stony Brook, Stony Brook, NY
11794--2100, USA}

{\small \noindent $^2$Instituto de F{\'\i}sica de Cantabria, Consejo Superior
de Investigaciones Cient{\'\i}ficas, $^3$Departamento de F{\'\i}sica Moderna,
Universidad de Cantabria, Facultad de Ciencias, 39005 Santander, SPAIN}

\vspace*{0.1in}
\hrule

\noindent This document presents supplementary information for an article
published in the 27 June 1996 edition of {\em Nature}, 381, 759--763.  The
supplementary information is available electronically from
ftp://ftp.\-ess.\-sunysb.\-edu/\-pub/\-hdf.

  Table 1 (17 pages) lists the identification number, WFPC2 chip number, WFPC2
pixel coordinates $x$ and $y$ of the trimmed images, J2000 Right Ascension and
Declination $\alpha$ and $\delta$, total magnitude $AB(8140)$,  relative fluxes
$f(3000)$, $f(4500)$, $f(6160)$, and $f(8140)$ (and uncertainties), and
maximum-likelihood redshift estimate $z$ of the 1,683 objects identified by the
analysis.  Uncertainties are given in parenthesis.  To obtain WFPC2 pixel
coordinates of the Version 2 drizzled images, add 190 to both $x$ and $y$.

  Figure 1 (38 pages) shows spectral energy distributions (left panels of each
pair) and redshift likelihood functions (right panels of each pair) of the
1,683 objects identified by the analysis.  For the spectral energy
distributions, vertical error bars show uncertainties, horizontal error bars
show bin sizes, and solid curves show best-fit model spectral energy
distributions.

  Figure 2 (1 page) shows optimally-added composite spectral energy
distributions of the objects identified by the analysis at redshift $z > 2.5$.
Vertical error bars show uncertainties and horizontal error bars show bin
sizes.  Vertical dashed lines indicate onset of the Lyman limit (shorter
wavelengths) and the \lya\ forest (longer wavelengths) at the midpoint of each
redshift bin.